\documentclass[journal]{IEEEtran}
\usepackage{graphicx, graphics, theorem, times, amsfonts, amsmath, amssymb, cite, caption}
\usepackage{tikz}
\usetikzlibrary{shapes,arrows}
\tikzstyle{phantom vertex} = [ ellipse, 
                               anchor = center, 
                               minimum height = 1*\unit, 
                               minimum width  = 1*\unit,
                               inner sep=0pt,
                               anchor=center]
\tikzstyle{red vertex}   = [black, fill = red!20,   phantom vertex, draw]
\tikzstyle{black vertex} = [black, fill = black!20, phantom vertex, draw]
\tikzstyle{blue vertex}  = [black, fill = blue!20,  phantom vertex, draw]
\tikzstyle{green vertex} = [black, fill = green!20,  phantom vertex, draw]
\tikzstyle{yellow vertex} = [black, fill = yellow!20,  phantom vertex, draw]
\tikzstyle{vertex}       = [draw, phantom vertex]

\tikzstyle{point} = [ellipse, inner sep=0pt, draw, fill=white, anchor = center,
                     minimum height = 0.05*\unit, minimum width  = 0.05*\unit]

\usepackage{pgfplots}
\usepackage{color}
\usepackage{needspace}

\newcommand{\myindentedparagraph}[1]{\needspace{1\baselineskip}\medskip \hangindent=11pt \hangafter=0 \noindent{\it #1.}}

\usepackage{hyperref}
\usepackage{multirow}
\usepackage{rotating}
\usepackage{cleveref, subcaption}
\usepackage{booktabs}
\usepackage{bm}
\input{mysymbol.sty}
\newtheorem{proposition}{\hspace{0pt}\bf Proposition}

\newtheorem{theorem}{\hspace{0pt}\bf Theorem}
\newtheorem{corollary}{\hspace{0pt}\bf Corollary}

\newtheorem{remark}{\hspace{0pt}\bf Remark}

\newcommand{\QED}{\hfill\ensuremath{\blacksquare}}

\begin{document}
\title{Metric Representations of Network Data}
\author{Santiago Segarra, Gunnar Carlsson, Facundo M\'emoli and Alejandro Ribeiro
\thanks{Work in this paper is supported by NSF CCF-1217963 and  AFOSR MURI FA9550-10-1-0567. A. Ribeiro and S. Segarra are with the Department of Electrical and Systems Engineering, University of Pennsylvania. G. Carlsson is with the Department of Mathematics, Stanford University. F. M\'emoli is with the Department of Mathematics and the Department of Computer Science and Engineering, Ohio State University. Email: ssegarra@seas.upenn.edu, gunnar@math.stanford.edu, memoli@math.osu.edu and aribeiro@seas.upenn.edu.}}

\maketitle

%%%%%%%%%%%%%%%%%%%%%%%%%%%%%%%%%%%%%%%%%%%%%%%%%%%%%%%%%%%%%%%
%
\begin{abstract}
Networks are structures that encode relationships between pairs of elements or nodes. However, there is no imposed connection between these relationships, i.e., the relationship between two nodes can be independent of every other one in the network, and need not be defined for \emph{every} possible pair of nodes. This is not true for metric spaces, where the triangle inequality imposes conditions that must be satisfied by triads of distances in the space, and these distances \emph{must} be defined for every pair of nodes. In this paper, we study how to project networks into $q$-metric spaces, a generalization of metric spaces that encompasses a larger class of structured representations. In order to do this, we encode as axioms two intuitively desirable properties of the mentioned projections. We show that there is only one way of projecting networks onto $q$-metric spaces satisfying these axioms. Moreover, for the special case of (regular) metric spaces, this method boils down to computing the shortest path between every node and, for the case of ultrametric spaces, it coincides with single linkage hierarchical clustering. Furthermore, we show that the projection method satisfies two properties of practical relevance: optimality, which enables its utilization for the efficient estimation of combinatorial optimization problems, and nestedness, which entails consistency of the structure induced when projecting onto different $q$-metric spaces. Finally, we illustrate how metric projections can be used to efficiently search networks aided by metric trees.
\end{abstract}

%%%%%%%%%%%%%%%%
%\begin{keywords}
%\end{keywords}

%!TEX root = metric_projections.tex

%%%%%%%%%%%%%%%%%%%%%%%%%%%%%%%%%%%%%%%%%%%%%%%%%%%%%%%%%%%%%%%
\section{Introduction}\label{sec_introduction}
%%%%%%%%%%%%%%%%%%%%%%%%%%%%%%%%%%%%%%%%%%%%%%%%%%%%%%%%%%%%%%%

Networks are data structures that encode relationships between elements and can be thought of as signals that, instead of having values associated with elements, have values associated with pairs of elements. As such, they play an important role in our current scientific understanding of problems in which relationships between elements are important. These problems include interactions between proteins or organisms in biology \cite{Bu03, Lieberman05}, individuals or institutions in sociology \cite{Newman06, Kleinberg99}, and neurons or regions in the brain \cite{Kleinberg06, Kempe04, Lynch96}. 

Despite their pervasive presence, tools to analyze networks and algorithms that exploit network data are not as well developed as tools and algorithms for processing of conventional signals. Although some of this lag can be attributed to different developmental stages, there is also the matter of the complexity of network data. To understand this latter point it is instructive to observe that particular cases of signals that encode relationships between elements are well understood and pose little challenge for analysis and algorithm design. E.g., a correlation matrix is a representation of the proximity between components of a random signal and a finite metric space defines distances between elements of a space. Both can be considered as particular cases of networks and the understanding of both is on par with the understanding of signals. Since correlation matrices and metric spaces have been studied for longer, the relative lag of network analysis can be again attributed to different developmental stages. However, the available evidence suggests otherwise.

Indeed, consider a problem of proximity search in which we are given a network and an element whose dissimilarity to different nodes of the network can be determined. We are asked to find the element that is least dissimilar. In an arbitrary network finding the least dissimilar node requires comparison against all nodes and incurs a complexity that is linear in the size of the network. In a metric space, however, the triangle inequality encodes a transitive notion of proximity. If two points are close to each other in a metric space and one of them is close to a third point, then the other one is also close to this third point. This characteristic can be exploited to design efficient search methods using metric trees whose complexity is logarithmic in the number of nodes \cite{Yianilos93, Uhlmann91, Traina00}. Likewise, many hard combinatorial problems on graphs are known to be approximable in metric spaces but not approximable in generic networks. The traveling salesman problem, for instance, is not approximable in generic graphs but it is approximable in polynomial time to within a factor of 3/2 in metric spaces \cite{christofides1976worst}. In either case, the advantage of the metric space is that the triangle inequality endows it with a structure that an arbitrary network lacks. It is this structure that makes network analysis and algorithm design tractable. 

If metric spaces are easier to handle than arbitrary networks, a possible route for network analysis is to design projection operators to map arbitrary networks into the subset of networks that represent metric spaces. This is the problem addressed in this paper.

\subsection{Related work and contributions}

The concept of an abstract metric space, introduced in the early 20th century \cite{Frechet06}, encompasses a wide variety of scientific and engineering constructions where the notion of distance is present. The main difference with weighted networks is that distances in metric spaces are governed by the triangle inequality whereas weights in networks need not satisfy any similar requirement. During the first half of the past century, metric spaces were regarded as mere presentations of underlying topological spaces and a lot of effort was put on the study of embedding general metric spaces into more familiar ones \cite{Blumenthal53}. However, in the late sixties, there was a partial shift in the focus of analysis, and the first formal studies of metric spaces \emph{as such} -- not seen as representations of some underlying topological space -- appeared \cite{Isbell64}, specifically in the field of category theory \cite{Barr90}. We leverage the fact that the fundamental understanding of metric spaces is more developed than that of networks in order to gain insight on the latter by projecting them onto the former and using analytical tools designed for the study of metric spaces.

The traditional way of mapping a generic dissimilarity function between pairs of points to a metric space is through multidimensional scaling (MDS) \cite{CoxCox08}. Different problem formulations give rise to the definition of different types of MDS where there is a basic distinction between metric MDS, where the input consists of quantitative similarities \cite{Torgerson52, Messick56}, and non-metric MDS where dissimilarities can be ordinal \cite{Kruskal64, Shepard62}. However, all these techniques have in common that the end goal is to facilitate visualization of the data \cite{Kruskal78}. Thus, MDS embeds the input dissimilarities into familiar and low-dimensional metric spaces such as $\reals^2$  or $\reals^3$. 

The advantages of metric embeddings exceed visualization. Since our interest exceeds data visualization, we propose an alternative procedure to map generic dissimilarity networks to metric spaces. In order to do this, we plan to follow an axiomatic approach, i.e., we encode as axioms desirable properties of the projections.

In Section~\ref{sec_networks_metric_spaces}, we formally introduce networks and $q$-metric spaces, the latter being an extension of metric spaces that enables the generalization of our framework to a variety of structured spaces. For instance, both the (regular) metric spaces and the ultrametric spaces \cite{jardine-sibson} can be seen as particular cases of $q$-metric spaces. In Section~\ref{sec_axioms_projection_transformation}, the two axioms of the proposed framework are introduced: a projection axiom -- networks that are already $q$-metric remain unchanged after projection -- and a transformation axiom -- smaller networks have smaller metric projections. We deem as admissible a projection method that satisfies these two axioms. The apparent weakness of these axioms contrast with the stringent theoretical consequences. More specifically, in Section~\ref{S:uniquness_metric_projections} we show that there is a unique admissible way of projecting networks onto $q$-metric spaces, that we call the \emph{canonical} $q$-metric projection. For the particular case of (regular) metric spaces, this implies that the shortest path between two nodes is the unique admissible distance between them. Moreover, when focusing on projections onto ultrametric spaces, single-linkage clustering \cite[Ch. 4]{clusteringref} is identified as the only admissible hierarchical clustering method. Furthermore, in Section~\ref{S:properties} we show that the axioms of Projection and Transformation confer two practical properties to canonical projections: optimality and nestedness. The former implies that the canonical projection of a network can be used to approximate the solution of combinatorial optimization problems on the network. Nestedness implies that the $q$-metric space obtained when projecting an arbitrary network is invariant to intermediate projections onto other $q'$-metric spaces with laxer structure. Finally, in Section~\ref{S:search}, we propose an efficient strategy for approximate search in networks by first projecting a general network onto a $q$-metric space and then leveraging this structure for search via the construction of a metric tree \cite{Yianilos93, Uhlmann91, Traina00}. We show that the parameter $q$ can be used to control the tradeoff between computational gain and search performance.

%!TEX root = metric_projections.tex

%%%%%%%%%%%%%%%%%%%%%%%%%%%%%%%%%%%%%%%%%%%%%%%%%%%%%%%%%%%%%%%
\section{Networks and $q$-metric spaces}\label{sec_networks_metric_spaces}
%%%%%%%%%%%%%%%%%%%%%%%%%%%%%%%%%%%%%%%%%%%%%%%%%%%%%%%%%%%%%%%

In the present paper we consider weighted and undirected graphs or \emph{networks}. Formally, we define a graph $G=(V, E, W)$ as a triplet formed by a finite set of $n$ nodes or vertices $V$, a set of edges $E \subset V \times V$ where $(x, y) \in E$ represents an edge from $x \in V$ to $y \in V$, and a map $W: E \to \reals_{++}$ from the set of edges to the strictly positive reals, representing weights $W(x,y)>0$ associated with each edge $(x,y)$. The weights are related to dissimilarities, i.e. the larger the weight the less similar the nodes are. The lack of direction in $G$ implies that if $(x, y) \in E$ then $(y, x) \in E$ and $W(x,y)=W(y,x)$. The graphs considered here are further assumed not to contain self-loops, i.e., $(x, x) \not\in E$ for all $x \in V$. Denote by $\ccalN$ the set of all networks. Networks in $\ccalN$ can have different node sets $V$, different edge sets $E$, or different weights $W$.

Given a set $X$, a metric $d: X \times X \to \reals_+$ is a function from the space of pairs of elements $X\times X$ to the non-negative reals satisfying the following three properties for every $x, y, z \in X$:
\begin{mylist}
\item[{\it Symmetry:}] $d(x, y)=d(y, x)$.
\item[{\it Identity:}] $d(x, y)=0$ if and only if $x=y$.
\item[{\it Triangle inequality:}] $d(x, y) \leq d(x, z) + d(z, y)$.
\end{mylist}
\noindent The ordered pair $M=(X, d)$ is said to be a \emph{finite metric space} \cite{Burago2001} and the space of all finite metric spaces is defined as $\ccalM$. 

In this paper we also consider tighter metric spaces that are termed \emph{$q$-metric spaces} and are parametrized by a constant $q \in [1, \infty]$. For strictly finite $q<\infty$, a $q$-metric space is a pair $M=(X, d)$ where the function $d$ satisfies the symmetry and identity properties but a q-triangle inequality in lieu of the regular triangle inequality. This $q$-triangle inequality is such that for all $x, y, z \in X$, it holds,
\begin{mylist}
\item[{\it q-triangle inequality:}] $d(x, y)^q \leq d(x, z)^q + d(z, y)^q$.
\end{mylist}
\noindent When $q=\infty$ we define an $\infty$-metric space as a pair $M=(X, d)$ that satisfies the symmetry and identity properties as well as the $\infty$-triangle inequality
\begin{mylist}
\item[{\it $\infty$-triangle inequality:}] $d(x, y) \leq \max[d(x, z) , d(z, y)]$, 
\end{mylist}
\noindent for all $x, y, z \in X$. Notice that the regular definition of metric space is recovered when $q=1$. When $q=\infty$, the $\infty$-triangle inequality is equivalent to the strong triangle inequality that characterizes ultrametric spaces -- which are equivalent representations of dendrograms, the outputs of hierarchical clustering methods \cite{clust-um, Carlsson2014}. Another instance of interest is that of 2-metric spaces, in which case all of the triangles in the space are acute angled triangles; see Section \ref{sec_2_metric_spaces}.

In this paper, we interpret $q$-metrics as particular cases of networks which we can do if we associate the network $M=(X, X\times X, d)$ with the $q$-metric space $M=(X,d)$. The space of all $q$-metric spaces is denoted here as $\ccalM_{q}$ and the space of metrics in general is denoted as $\ccalM\equiv\ccalM_1$. Notice that for all $q > q' > 1$ we must have
\begin{equation}\label{eqn_metric_inclusions}
   \ccalM_{q} \subset \ccalM_{q'} 
              \subset \ccalM_{1} \equiv \ccalM 
              \subset \ccalN.
\end{equation}
A closely related definition is that of a norm \cite{Burago2001}. In this case, for a given vector space $Y$, a norm $\| \cdot \|$ is a function $\| \cdot \| : Y \to \reals_+$ from $Y$ to the non-negative reals such that, for all vectors $v, w \in Y$ and scalar constant $\beta$, it satisfies:

\begin{mylist}
\item[{\it Positiveness:}] $\|v\| \geq 0$ with equality if and only if $v = \vec{0}$.
\item[{\it Positive homogeneity:}] $\|\beta \,w \| = |\beta| \,\| w \|$.
\item[{\it Subadditivity:}] $\|v + w\| \leq \|v\| + \|w\|$.
\end{mylist}
A commonly used family of norms for vectors in $\reals^l$ are the $p$-norms $\| \cdot \|_p$ for real $p \in [1, \infty]$, where the norm of a vector $v = [v_1, \ldots, v_l]$ is given by
\begin{equation}\label{E:def_p_norms}
\| v \|_p =
\begin{cases}
\left( \sum_{i=1}^l |v_i|^p \right)^{1/p} \quad\,\, \text{for} \,\, p<\infty,\\
\max_i |v_i| \qquad\qquad \,\,\,\,  \text{for} \,\, p=\infty.
\end{cases}
\end{equation}

In the definition of metric projections, the concepts of path and path length are important. Given a network $G=(V, E, W)$ and $x, x' \in V$, a path $P_{xx'}$ is an \emph{ordered} sequence of nodes in $V$, 
\begin{equation}\label{eqn_definition_path}
   P_{xx'}=[x=x_0, x_1, \ldots , x_{l-1}, x_l=x'],
\end{equation}
which starts at $x$ and finishes at $x'$ and $e_i=(x_i, x_{i+1}) \in E$ for $i=0, \ldots, l-1$. We say that $P_{xx'}$ links or connects $x$ to $x'$. 
The links $e_i$ of a path are the edges connecting consecutive nodes of the path in the direction given by the path. In the present paper we refer exclusively to connected graphs, i.e., graphs where there exist paths $P_{xx'}$ for all $x, x' \in V$. Given two paths $P_{xx'}=[x=x_0, x_1, ... , x_l=x']$ and $P_{x'x''}=[x'=x'_0, x'_1, ... , x'_{l'}=x'']$ such that the end point $x'$ of the first one coincides with the starting point of the second one, we define the \emph{concatenated path} $P_{xx'} \oplus P_{x'x''}$ as
\begin{align}\label{eqn_definition_concatenation}
P_{xx'} \oplus P_{x'x''} = [x=x_0,\ldots , x_l=x'=x'_0,\ldots , x'_{l'}=x'']. 
\end{align}
For a given norm $\| \cdot \|$, we define the \emph{length} of a given path $P_{xx'}=[x=x_0,\ldots, x_l=x']$ as $h_{\| \cdot \|}(P_{xx'}) = \| [W(x_0,x_1), \ldots, W(x_{l-1},x_l)] \|$, i.e., the norm of the vector that consists of the weights associated to the links in the path. For notational consistency, we utilize the convention that  $h_{\| \cdot \|}([x, x]) = 0$ for all $x \in V$. We define the minimum length function $s_{\| \cdot \|}: V \times V \to \reals_+$ where the minimum length $s_{\| \cdot \|}$ between nodes $x, x' \in V$ is given by
\begin{equation}\label{eqn_shortest_path}
s_{\| \cdot \|}(x,x') = \min_{P_{xx'}} \,\, h_{\| \cdot \|}(P_{xx'}).
\end{equation}
The connectedness of $G$ ensures that $s_{\| \cdot \|}(x,x')$ is well-defined for every pair of nodes $x, x' \in V$. 
%In particular, when the norm of interest is $\| \cdot \|_1$ [cf. \eqref{E:def_p_norms}], expression \eqref{eqn_shortest_path} reduced to the regular shortest path
%
%\begin{equation}\label{eqn_shortest_path_2}
%s_{\| \cdot \|_1}(x,x') = \min_{P_{xx'}} \sum_{i=0}^{l-1} w_{x_ix_{i+1}},
%\end{equation}
%
%i.e., the minimum length across paths connecting $x$ and $x'$. Similarly, if we specialize \eqref{eqn_shortest_path} for the $\infty$-norm, we minimize the maximum dissimilarity encountered when traveling between two nodes, i.e.
%
%\begin{equation}\label{eqn_shortest_path_3}
%s_{\| \cdot \|_\infty}(x,x') = \min_{P_{xx'}} \,\, \max_{i} \,\, w_{x_ix_{i+1}},
%\end{equation}
%
%Another relevant property of a network $G=(V, E, W)$ is the \emph{separation} $\sep(G)$ which we define as the minimum weight in the graph, 
%
%\begin{equation}\label{eqn_def_separation_network}
%   \sep(G) = \min_{(x, y) \in E} w_{xy}.
%\end{equation}
%

In the present paper we study the design of $q$-metric projections $\ccalP_q$ with the objective of representing networks as $q$-metric spaces. Formally, for all node sets $V$ we define a $q$-metric projection $\ccalP_q: \ccalN \to \ccalM_{q}$ as a map that projects every network onto a $q$-metric space while preserving $V$. We say that two $q$-metric projections $\ccalP_q$ and $\ccalP'_q$ are \emph{equivalent}, and we write $\ccalP_q \equiv \ccalP'_q$, if and only if $\ccalP_q(G) = \ccalP'_q(G)$, for all $G \in \ccalN$.

%!TEX root = metric_projections.tex

%%%%%%%%%%%%%%%%%%%%%%%%%%%%%%%%%%%%%%%%%%%%%%%%%%%%%%%%%%%%%%%
\section{Axioms of Projection and Transformation}\label{sec_axioms_projection_transformation}
%%%%%%%%%%%%%%%%%%%%%%%%%%%%%%%%%%%%%%%%%%%%%%%%%%%%%%%%%%%%%%%

The broad definition of $\ccalP_q$ as a node-preserving and $q$-metric inducing map presented in Section~\ref{sec_networks_metric_spaces} admits maps of undesirable behavior. E.g., we may define the map $\ccalP^1_q$ such that for any $G=(V, E, W)$ it outputs the space $(V, d) = \ccalP^1_q(G)$ where $d$ is defined as $d(x, x')=1$ and $d(x, x)=0$ for all $x \neq x' \in V$. It is immediate to see that the output space is $q$-metric for all $q$, since it is $\infty$-metric or ultrametric. However, the defined $q$-metric $d$ completely ignores the edge structure in $E$ and the edge weights in $W$, which is undesirable. In order to discard unreasonable projections like $\ccalP^1_q$, we encode in the form of the axioms of Projection and Transformation, desirable properties that a map $\ccalP_q$ should satisfy.

The space of $q$-metric spaces $\ccalM_{q}$ is a subspace of the space of all networks $\ccalN$; see Section~\ref{sec_networks_metric_spaces}. Formally, we associate every $q$-metric space $(V, d) \in \ccalM_{q}$ to the complete network $(V, E, W_d)$ where the edge set contains every possible edge except self-loops, i.e. $(x, y) \in E$ for all $x \neq y \in V$. Furthermore, the edge weights $W_d$ are given by the $q$-metric function $d$, i.e., $W_d(x, y) = d(x, y)$. Throughout the paper, we represent $q$-metric spaces as $(V, d)$ or its network equivalent $(V, E, W_d)$ interchangeably. Based on this observation, we define the following axiom:

%%%   F   I   G   U   R   E   %%%%%%%%%%%%%%%%%%%%%%%%%%%%%%%%%%%
%
\begin{figure}
\centering
\vspace{-0pt}
\def \thisplotscale {0.5}
\def \unit {\thisplotscale cm}

\begin{tikzpicture}[-stealth, shorten >=2, x = 1*\unit, y=0.9*\unit, font=\footnotesize]

    % Draw space blurbs
    \path (0,0) node [draw, fill = blue!10, ellipse, 
                     minimum width = 4*\unit, 
                     minimum height = 6*\unit] (networks) {} 
          ++ (0,2) node {$\ccalN$};

   % \path (0,-1) node [draw, fill = blue!20, ellipse, 
     %               minimum width = 3*\unit, 
       %              minimum height = 3*\unit] (ultrametrics subset) {$\ccalM_q$};
	
	%    \path (9,-1) node [draw, fill = blue!20, ellipse, 
           %          minimum width = 3*\unit, 
             %        minimum height = 3*\unit] (ultrametrics) {$\ccalM_q$};
	
	\draw [fill = blue!20] plot [smooth cycle, tension=1] coordinates {(-1.5, -1) (1,-2.5) (0.5,-1) (1,0.5)};
    	\node at (-0.5, -1) {$\ccalM_q$};

	\draw [fill = blue!20] plot [smooth cycle, tension=1] coordinates {(-1.5+9, -1) (1+9,-2.5) (0.5+9,-1) (1+9,0.5)};
    	\node at (-0.5+9, -1) {$\ccalM_q$};

    % Draw arrows
    \path[-stealth] (networks)
                    edge [thick, bend left, above, 
                    pos=0.40, 
                    out=50, 
                    in = 130] node {$\ccalP_q$} 
                    (-1.5+9+1, -0);

   \path[-stealth]  (0.5,-1)
                       edge [above, thick] node {$\ccalP_q$} 
                    (-1.5+9, -1);

\end{tikzpicture}
\caption{Axiom of Projection. The $q$-metric space $\ccalM_q$ is an invariant set of the projection map $\ccalP_q$.}
\label{fig:axiom_of_projection}
\end{figure}
%%%%%%%%%%%%%%%%%%%%%%%%%%%%%%%%%%%%%%%%%%%%%%%%

\myindentedparagraph{(A1) Axiom of Projection} The $q$-metric space $M \in \ccalM_{q}$ is a fixed point of the projection map $\ccalP_q$, i.e. $\ccalP_q(M) = M$.

\medskip\noindent Given that our goal is the design of maps that transform general networks into more structured $q$-metric spaces, if we already have a $q$-metric space there is no justification to change it; see Fig.~\ref{fig:axiom_of_projection}. This concept is captured in axiom (A1) where we define $\ccalM_{q}$ as the fixed set of $\ccalP_q$. Equivalently, we say that the map $\ccalP_q$ restricted to $\ccalM_{q}$ is the identity map. It is immediate that axiom (A1) implies idempotency of $\ccalP_q$, which is a requirement of projection maps, hence, its denomination as Axiom of Projection.

The second restriction on the space of allowable maps $\ccalP_q$ formalizes our expectations for the behavior of $\ccalP_q$ when confronted with a transformation of the underlying node space $V$, edge space $E$, and edge weights $W$. Consider networks $G=(V,E, W)$ and $G'=(V',E', W')$, and denote by $M=\ccalP_q(G) = (V, d)$ and $M'=\ccalP_q(G') = (V', d')$ the corresponding $q$-metric space outputs when applying $\ccalP_q$ to $G$ and $G'$, respectively. 
If we can map all the nodes of the network $G$ into the nodes of the network $G'$ in such a way that no edge weight is increased, then we expect the points in $M'$ to be closer to each other than those in $M$. In order to formalize this notion, we introduce the following concept: given two networks $G=(V, E, W)$ and $G'=(V', E', W')$, the \emph{injective} map $\phi : V \to V'$ is called \emph{dissimilarity-reducing} map if it holds that $(\phi(x),\phi(x')) \in E'$ and $W(x,x') \geq W(\phi(x),\phi(x'))$ for all $(x, x') \in E$; see Fig.~\ref{fig:dissimilarity_reducing_map}. A dissimilarity-reducing map matches every edge in $G$ with an edge in $G'$ of less or equal weight. 

Intuitively, if we look at any path $P_{xx'}$ between nodes $x$ and $x'$ in $G$, the existence of the dissimilarity-reducing map $\phi$ ensures that there is an associated path $P_{\phi(x)\phi(x')}$ in $G'$ between the nodes $\phi(x)$ and $\phi(x')$ such that the weight of every link in this second path is not greater than the corresponding links in the first one. On top of this, there might exist additional paths between $\phi(x)$ and $\phi(x')$ in $G'$ that are not the image under $\phi$ of any path between $x$ and $x'$ in $G$. Thus, it is expected for nodes $\phi(x)$ and $\phi(x')$ to be closer to each other in the output $q$-metric spaces.

%%%%%%%%%%%%%%%   F   I   G   U   R    E    %%%%%%%%%%%%%%%%%%%%%
\begin{figure}
\centering
\def \thisplotscale {0.7}
\def \unit {\thisplotscale cm}

{\small
\begin{tikzpicture}[-stealth, shorten >=2, x = 1*\unit, y=0.9*\unit]

    % Draw bigger network
    \node [blue vertex, minimum height=0.6cm, minimum width=0.6cm] at (   0,  3.4) (1) {$x_1$};
    \node [blue vertex, minimum height=0.6cm, minimum width=0.6cm] at ( 0, -0.6) (2) {$x_2$};    
    \node [blue vertex, minimum height=0.6cm, minimum width=0.6cm] at (-2.5, -0.6) (3) {$x_3$};

   % \path [stealth-stealth] (1) edge [right] node {{$1$}} (2);	
    \path [stealth-stealth]  (2) edge [below] node {{$1$}} (3);
    \path [stealth-stealth]  (3) edge [above left] node  {{$2$}} (1);    	
    
    % Draw smaller network
    \node [blue vertex, minimum height=0.6cm, minimum width=0.6cm] at (3.0,1.0) (1p) {$y_3$};
    \node [blue vertex, minimum height=0.6cm, minimum width=0.6cm] at (5.0,3.0) (2p) {$y_1$};    
    \node [blue vertex, minimum height=0.6cm, minimum width=0.6cm] at (5.0,-1.0) (3p) {$y_2$};
    \node [blue vertex, minimum height=0.6cm, minimum width=0.6cm] at (7.0,1.0) (4p) {$y_4$};

    \path [stealth-stealth]  (1p) edge [above left] node {{ $1$}} (2p);	
    \path [stealth-stealth]  (2p) edge [right] node {{$2$}} (3p);
    \path [stealth-stealth]  (3p) edge [above right]  node {{$1$}} (1p);    	
    \path [stealth-stealth]  (2p) edge [above right]  node {{$3$}} (4p);

    % Connect nodes with transformation \phi  	
    \path (1) edge [bend left, above, red, very thick, pos=0.5] node {$\bbphi$} (2p);	
    \path (2) edge [bend left, above, red, very thick, pos=0.3] node {$\bbphi$} (3p);	    
    \path (3) edge [bend left, above, red, very thick, pos=0.5] node {$\bbphi$} (1p);

     % Write graph names
     \node at (0-2,  3.4) (G_1) {${\bf G}$};
     \node at (0+7,  3.4) (G_2) {${\bf G'}$};

\end{tikzpicture}
}
\caption{Dissimilarity-reducing map. The injective map $\phi$ takes every edge in network $G$ to an edge in network $G'$ of less or equal weight.}
\label{fig:dissimilarity_reducing_map}
\end{figure}
%%%%%%%%%%%%%%%%%%%%%%%%%%%%%%%%%%%%%%%%%%%%%%%%

The Axiom of Transformation that we introduce next is a formal statement of the intuition described above:

\myindentedparagraph{(A2) Axiom of Transformation} Consider any two networks $G=(V,E,W)$ and $G'=(V', E', W')$ and any dissimilarity-reducing map $\phi: V \to V'$. Then,  for all $x, x' \in V$, the output $q$-metric spaces $(V,d)=\ccalP_q(G)$ and $(V',d')=\ccalP_q(G')$ satisfy 
\begin{equation}\label{eqn_dissimilarity_reducing_q_metric}
    d(x,x') \geq d'(\phi(x),\phi(x')).
\end{equation} 

\medskip\noindent  We say that a projection $\ccalP_q$ is \emph{admissible} if it satisfies axioms (A1) and (A2).
In the next section, we study the space of admissible projections $\ccalP_q$ for every $q$.

\begin{remark}[Related axiomatic constructions]\label{R:related_axiomatic_constructions}\normalfont
In \cite{clust-um,CarlssonMemoli10} the authors propose three axioms to study hierarchical clustering of metric spaces. Given the relation between hierarchical clustering and ultrametrics \cite{clust-um, Carlsson2014}, such a problem can be reformulated in terms of our notation as the definition of maps from $\ccalM_{1}$ to $\ccalM_{\infty}$. In our current work, the domain set is $\ccalN$ (which is a superset of $\ccalM_{1}$) and we use a unified framework to study projections onto all $\ccalM_{q}$, of which a particular case is $q=\infty$. The axioms of Projection and Transformation are related to two of the axioms in \cite{clust-um,CarlssonMemoli10}. In Section~\ref{Ss:ultrametric_spaces} we show that, using our more general framework, we recover the unicity result in \cite{clust-um,CarlssonMemoli10} with less stringent axioms. In \cite{Carlsson2014}, the authors use a similar axiomatic framework to study the projection of asymmetric networks onto $\ccalM_{\infty}$. The network model in \cite{Carlsson2014} is different to the one in the current paper. It is more restrictive in the sense that complete graphs are assumed but, at the same time, more general since directed graphs are allowed.
Extensions of the axiomatic framework here presented for the projection of \emph{directed} networks onto general $q$-metric spaces are left as future work.
\end{remark}

%!TEX root = metric_projections.tex

%%%%%%%%%%%%%%%%%%%%%%%%%%%%%%%%%%%%%%%%%%%%%%%%%%%%%%%%%%%%%%%
\section{Uniqueness of Metric Projections}\label{S:uniquness_metric_projections}
%%%%%%%%%%%%%%%%%%%%%%%%%%%%%%%%%%%%%%%%%%%%%%%%%%%%%%%%%%%%%%%

After posing the axioms of Projection and Transformation, we first seek to answer if \emph{any} map $\ccalP_q$ satisfies them. In this direction, given a graph $G=(V, E, W)$, for each $q$ we define the \emph{canonical} $q$-metric projection $\ccalP_q^*$ with output $(V, d_q^*)= \ccalP_q^*(G)$ where the $q$-metric $d_q^*$ between points $x$ and $x'$ is given by
\begin{equation}\label{E:def_canonical_q_projection}
d_q^*(x, x') = s_{\| \cdot \|_q}(x,x') = \min_{P_{xx'}} h_{\| \cdot \|_q}(P_{xx'}).
\end{equation}
In \eqref{E:def_canonical_q_projection}, to find the distance between two points, we look for the path that links these nodes while minimizing its $q$-norm, as defined in \eqref{E:def_p_norms}.

For the method $\ccalP_q^*$ to be a properly defined $q$-metric projection for all $q$, we need to establish that $(V, d_q^*)$ is a valid $q$-metric space. Furthermore, it can also be shown that $\ccalP_q^*$ satisfies axioms (A1)-(A2). We prove both assertions in the following proposition.

%%%   P R O P O S I T I O N %%%%%%%%%%%%%%%%%%%%%%%%%%%%%%%%%%%%%%%%%
\begin{proposition}\label{prop_canonical_projection_axioms}
The canonical $q$-metric projection map $\ccalP_q^*$ is valid and admissible. I.e., $d_q^*$ defined by \eqref{E:def_canonical_q_projection} is a $q$-metric for all graphs $G$ and $\ccalP_q^*$ satisfies the axioms of Projection (A1) and Transformation (A2).
\end{proposition}
%%%%%%%%%%%%%%%%%%%%%%%%%%%%%%%%%%%%%%%%%
\begin{myproof}
We first prove that $d_q^*$ is indeed a $q$-metric on the node space $V$. That $d_q^*(x, x') = d_q^*(x', x)$ follows from combining the facts that the original graph $G$ is symmetric and that the norms $\|\cdot\|_q$ are symmetric in the elements of the vectors for all $q$. Moreover, that $d_q^*(x,x')=0$ if and only if $x=x'$ is a consequence of the positiveness property of the $q$-norms. To verify that the q-triangle inequality holds, let $P_{xx'}$ and $P_{x'x''}$ be paths that achieve the minimum in \eqref{E:def_canonical_q_projection} for $d_q^*(x,x')$ and $d_q^*(x',x'')$, respectively. Then, from definition \eqref{E:def_canonical_q_projection} it follows that
\begin{align}\label{E:proof_prop_canonical_projection_axioms_010}
{d_q^*(x, x'')}^q &= \min_{P_{xx''}} {h_{\| \cdot \|_q}(P_{xx''})}^q \leq {h_{\| \cdot \|_q}(P_{xx'} \oplus P_{x'x''})}^q \nonumber \\
&= {h_{\| \cdot \|_q}(P_{xx'})}^q + {h_{\| \cdot \|_q}(P_{x'x''})}^q \nonumber \\
& = {d_q^*(x,x')}^q + {d_q^*(x',x'')}^q,
\end{align}
where the inequality follows from the fact that the concatenated path $P_{xx'} \oplus P_{x'x''}$ is \emph{a particular} path between $x$ and $x''$ while the definition of $d_q^*(x, x'')$ minimizes the norm across all such paths.

To see that the Axiom of Projection (A1) is satisfied, pick an arbitrary $q$-metric space $M=(V, d) \in \ccalM_q$ and denote by $(V, d_q^*) = \ccalP_q^*(M)$ the output of applying the canonical $q$-metric projection to $M$. For an arbitrary pair of nodes $x, x' \in V$, we have that
\begin{equation}\label{E:proof_prop_canonical_projection_axioms_020}
d_q^*(x, x') = \min_{P_{xx'}} h_{\| \cdot \|_q}(P_{xx'}) \leq h_{\| \cdot \|_q}([x, x']) = d(x, x'),
\end{equation}
for all $q$, where the inequality comes from specializing the path $P_{xx'}$ to the path $[x, x']$ with just one link from $x$ to $x'$. Moreover, if we denote by $P^*_{xx'} = [x = x_0, x_1, \ldots, x_l=x']$ the path achieving the minimum in \eqref{E:proof_prop_canonical_projection_axioms_020}, then we may leverage the fact that $d$ satisfies the q-triangle inequality to write
\begin{equation}\label{E:proof_prop_canonical_projection_axioms_030}
d(x, x') \leq \left(\sum_{i=0}^{l-1}  d(x_i, x_{i+1})^q\right)^{1/q} \!\!\!\!= h_{\| \cdot \|_q}(P^*_{xx'})  = d_q^*(x, x').
\end{equation}
Upon substituting \eqref{E:proof_prop_canonical_projection_axioms_030} into \eqref{E:proof_prop_canonical_projection_axioms_020}, we obtain that all the inequalities are, in fact, equalities, implying that $d_q^*(x, x') = d(x, x')$. Since nodes $x, x'$ were chosen arbitrarily, it must be that $d \equiv d_q^*$ which implies that $\ccalP_q^*(M) = M$, as wanted.

To show fulfillment of axiom (A2), consider two networks $G=(V, E, W)$ and $G' = (V', E', W')$ and a dissimilarity-reducing map $\phi:V \to V'$. Let $(V, d_q)=\ccalP_q^*(G)$ and $(V', d'_q)=\ccalP_q^*(G')$ be the outputs of applying the canonical projection to networks $G$ and $G'$, respectively.
For an arbitrary pair of nodes $x, x' \in V$, denote by $P^*_{xx'}=[x=x_0,\ldots, x_l=x']$ a path that achieves the minimum in \eqref{E:def_canonical_q_projection} so as to write
\begin{equation}\label{E:proof_prop_canonical_projection_axioms_040}
    d_q(x,x') = h_{\| \cdot \|_q}(P^*_{xx'}).
\end{equation}
Consider the transformed path $P^*_{\phi(x)\phi(x')}=[\phi(x)=\phi(x_0),\ldots, \phi(x_l)=\phi(x')]$ in the space $V'$. Since the transformation $\phi$ does not increase dissimilarities, we have that for all links in this path $W'(\phi(x_i),\phi(x_{i+1})) \leq W(x_i,x_{i+1})$. Combining this observation with \eqref{E:proof_prop_canonical_projection_axioms_040} we obtain,
\begin{align}\label{E:proof_prop_canonical_projection_axioms_050}
    h_{\| \cdot \|_q}(P^*_{\phi(x)\phi(x')}) \leq d_q(x,x').
\end{align}
Further note that $P_{\phi(x)\phi(x')}$ is a particular path joining $\phi(x)$ and $\phi(x')$ whereas the metric $d'_q$ is given by the minimum across all such paths. Therefore,
\begin{equation}\label{E:proof_prop_canonical_projection_axioms_060}
    d'_q(\phi(x),\phi(x')) \leq h_{\| \cdot \|_q}(P^*_{\phi(x)\phi(x')}).
   \end{equation}
Upon replacing \eqref{E:proof_prop_canonical_projection_axioms_050} into \eqref{E:proof_prop_canonical_projection_axioms_060}, it follows that $d'_q(\phi(x),\phi(x')) \leq d_q(x,x')$, as required by the Axiom of Transformation.
\end{myproof}
%%%%%%%%%%%%%%%%%%%%%%%%%%%%%%%%%%%%%%%%%

Given that we have shown that the canonical $q$-metric projection $\ccalP_q^*$ satisfies axioms (A1)-(A2), two questions arise: i) are there other projections satisfying (A1)-(A2)?; and ii) is the projection $\ccalP_q^*$ special in any sense? Both questions are answered by the following uniqueness theorem.

%%%   T H E O R E M  %%%%%%%%%%%%%%%%%%%%%%%%%%%%%%%%%%%%%%%%%
\begin{theorem}\label{theo_unicity_q_metric}
Let $\ccalP_q:\ccalN\to\ccalM_q$ be a $q$-metric projection, and $\ccalP_q^*$ be the canonical projection with output $q$-metric as defined in \eqref{E:def_canonical_q_projection}. If $\ccalP_q$ satisfies the axioms of Projection (A1) and Transformation (A2) then $\ccalP_q\equiv\ccalP_q^*$, for all $q$.
\end{theorem}
%%%%%%%%%%%%%%%%%%%%%%%%%%%%%%%%%%%%%%%%%
\begin{myproof}
Given an arbitrary network $G=(V, E, W)$ and a fixed $q$, denote by $(V, d) = \ccalP_q(G)$ and $(V, d^*) = \ccalP_q^*(G)$ the output $q$-metric spaces when applying a generic admissible $q$-metric projection and the canonical $q$-metric projection, respectively. We will show that
\begin{equation}\label{E:proof_theo_unicity_q_metric_010}
d^*(x, x') \leq d(x, x') \leq d^*(x, x'),
\end{equation}
for all $x, x'\in V$. Given that $G$ was chosen arbitrarily, this implies that $\ccalP_q\equiv\ccalP_q^*$, as wanted.

We begin by showing that $d(x, x') \leq d^*(x, x')$ for all $x, x' \in V$. Consider an arbitrary pair of points $x$ and $x'$  and let $P_{xx'}=[x=x_0,\ldots, x_l=x']$ be a path achieving the minimum in \eqref{E:def_canonical_q_projection} so that we can write
\begin{equation}\label{E:proof_theo_unicity_q_metric_020}
   d^*(x, x') =  \left(\sum_{i=0}^{l-1}  W(x_i,x_{i+1})^q\right)^{1/q}.
\end{equation}
Focus now on a series of two-node networks $G_i = (V_i, E_i, W_i)$ for $i=0, \ldots, l-1$, such that $V_i = \{z, z'\}$ and $E_i=\{(z,z'), (z',z)\}$ for all $i$ but with different weights given by $W_i(z,z') = W_i(z',z) = W(x_i,x_{i+1})$. Since every network $G_i$ is already a $q$-metric -- in fact, any two-node network is a valid $q$-metric for all $q$ -- and the method $\ccalP_q$ satisfies the Axiom of Projection (A1), if we define $(\{z, z'\}, d_i) = \ccalP_q(G_i)$ we must have that $d_i(z,z') = W(x_i, x_{i+1})$, i.e., every graph $G_i$ is a fixed point of the map $\ccalP_q$. 

Consider transformations $\phi_i:\{z,z'\} \to V$ given by $\phi_i(z)=x_i$, $\phi_i(z')=x_{i+1}$ so as to map $z$ and $z'$ in $G_i$ to subsequent points in the path $P_{xx'}$ used in \eqref{E:proof_theo_unicity_q_metric_020}. This implies that maps $\phi_i$ are dissimilarity-reducing since they are injective and the only edge in $G_i$ is mapped to an edge of the exact same weight in $G$ for all $i$. Thus, it follows from the Axiom of Transformation (A2) that
\begin{align}\label{E:proof_theo_unicity_q_metric_030}
   d(\phi_i(z),\phi_i(z')) = d(x_i, x_{i+1}) \leq d_i(z,z') = W(x_i,x_{i+1}).
\end{align}
To complete the proof we use the fact that since $d$ is a $q$-metric and $P_{xx'}=[x=x_0,\ldots, x_l=x']$ is a path joining $x$ and $x'$, the q-triangle inequality dictates that
\begin{align}\label{E:proof_theo_unicity_q_metric_040}
   d(x,x') \leq \left(\sum_{i=0}^{l-1}  d(x_i, x_{i+1})^q\right)^{1/q} \!\!\!\!\! \leq \left(\sum_{i=0}^{l-1}  W(x_i,x_{i+1})^q\right)^{1/q}\!\!\!\!\!\!,
\end{align}
where we used \eqref{E:proof_theo_unicity_q_metric_030} for the second inequality. The proof that $d(x, x') \leq d^*(x, x')$ follows from substituting \eqref{E:proof_theo_unicity_q_metric_020} into \eqref{E:proof_theo_unicity_q_metric_040}.

We now show that $d(x, x') \geq d^*(x, x')$ for all $x, x' \in V$. To do this, first notice that for an arbitrary pair of points $x$ and $x'$, if the edge $(x, x') \in E$ then we have that
\begin{align}\label{E:proof_theo_unicity_q_metric_050}
   d^*(x,x') =  \min_{P_{xx'}} h_{\| \cdot \|_q}(P_{xx'}) \leq W(x,x'),
\end{align}
where the inequality comes from considering the particular path $P_{xx'}$ with only two points $[x, x']$. Hence, the identity map $\phi: V \to V$ such that $\phi(x) = x$ for all $x \in V$ is a dissimilarity reducing map from $G$ to $(V, d^*)$, since it is injective and every existing edge in $G$ is mapped to an edge with smaller or equal weight. Consequently, we can build the diagram of relations between spaces depicted in Fig.~\ref{fig:relation_maps_proof_unicity}. The top (blue) and left (red) maps in the figure are given by the definitions at the beginning of this proof while the relation on the right (green) is a consequence of the axiom (A1). Since the aforementioned identity map $\phi$ is dissimilarity reducing, we can use the fact that $\ccalP_q$ satisfies axiom (A2) to say that
\begin{align}\label{E:proof_theo_unicity_q_metric_060}
   d(x,x') \geq d^*(\phi(x),\phi(x')) = d^*(x,x'),
\end{align}
for all $x, x' \in V$, concluding the proof.
\end{myproof}

%%%%%%%%%%%%%%%   F   I   G   U   R    E    %%%%%%%%%%%%%%%%%%%%%
\begin{figure}
\centering
\def \thisplotscale {0.7}
\def \unit {\thisplotscale cm}

{\large
\begin{tikzpicture}[-stealth, shorten >=2, x = 1*\unit, y=0.9*\unit]

        \node at (0,0) (1) {${\bf G}$};
        \node at (4,0) (2) {${\bf (V, d^*)}$};
        \node at (0,-3) (3) {${\bf (V, d)}$};
        \node at (4,-3) (4) {${\bf (V, d^*)}$};
        
        \path (1) edge [very thick, above, blue] node {$\ccalP_q^*$} (2);	
        \path (1) edge [very thick, left, red] node {$\ccalP_q$} (3);	
	\path (2) edge [very thick, right, black!50!green] node {$\ccalP_q$} (4);	

\end{tikzpicture}
}
\caption{Diagram of maps between spaces for the proof of Theorem~\ref{theo_unicity_q_metric}. }
\label{fig:relation_maps_proof_unicity}
\end{figure}
%%%%%%%%%%%%%%%%%%%%%%%%%%%%%%%%%%%%%%%%%%%%%%%%

According to Theorem~\ref{theo_unicity_q_metric}, for each $q$ there is one and only one projection map from the space of networks onto $\ccalM_q$ that satisfies the axioms of Projection and Transformation, and this map is $\ccalP_q^*$. Any other conceivable map into $\ccalM_q$ must violate at least one of the axioms. E.g., if we have $q>q'$ then $\ccalP_q^*$ can be viewed as a map into $\ccalM_{q'}$ since $\ccalM_{q} \subset \ccalM_{q'}$. However, map $\ccalP_q^*$ violates axiom (A1) when viewed as a map into $\ccalM_{q'}$, since any $q'$-metric space $M \in \ccalM_{q'}$ that is \emph{not} in $\ccalM_{q}$ would not be a fixed point of $\ccalP_q^*$.

\subsection{Metric spaces}\label{Ss:metric_spaces}

For an arbitrary network $G=(V, E, W)$, we may particularize our analysis to the case of (regular) metric spaces, i.e. $q=1$. In this case, the canonical projection $\ccalP_1^*$ outputs the metric space $(V, d^*_1) = \ccalP_1^*(G)$ where $d^*_1$ is given by [cf. \eqref{E:def_canonical_q_projection}]
\begin{equation}\label{E:canonical_projection_q_1}
d^*_1(x, x') = \min_{P_{xx'}} \sum_{i=0}^{l-1} W(x_i, x_{i+1}),
\end{equation}
for all $x, x' \in V$. Equivalently, $\ccalP_1^*$ sets the distance between two nodes to the length of the shortest path between them. 

By specializing Theorem~\ref{theo_unicity_q_metric} to the case $q=1$, we obtain the following corollary.
%
%%%  C O R O L L A RY  %%%%%%%%%%%%%%%%%%%%%%%%%%%%%%%%%%%%%%%%%
\begin{corollary}\label{cor_unicity_1_metric}
The shortest path between every pair of nodes is the only admissible metric in networks, where admissibility is given by axioms (A1)-(A2).
\end{corollary}
%%%%%%%%%%%%%%%%%%%%%%%%%%%%%%%%%%%%%%%%%

From the comparison of networks \cite{Rubinov20101059} to the determination of node importance \cite{Freeman77, segarra2014stability}, shortest paths constitute a basic feature of network analysis and their application is ubiquitous. Moreover, efficient algorithms for the computation of every shortest path in a network exist \cite{Floyd62, Warshall62}. In Section~\ref{Ss:optimality}, we discuss the utility of projecting graphs onto metrics for the approximation of otherwise NP-hard graph theoretical problems that have guaranteed error for metric data \cite{FernandezdelaVega04bisection, Indyk99sublinear}.

\subsection{Ultrametric spaces}\label{Ss:ultrametric_spaces}

The equivalence between dendrograms and ultrametric spaces is a well-established fact \cite{clust-um, Carlsson2014}. Hence, hierarchical clustering of networks \cite{lance67general, clusteringref, ZhaoKarypis05}, i.e., the mapping of networks onto dendrograms, can be posed as a projection problem of networks onto ultrametric spaces. When setting $q = \infty$, the canonical projection induces the ultrametric $d^*_\infty$ given by [cf. \eqref{E:def_canonical_q_projection}]
\begin{equation}\label{E:canonical_projection_q_infty}
d^*_\infty(x, x') = \min_{P_{xx'}} \max_i W(x_i, x_{i+1}),
\end{equation}
for all $x, x' \in V$. Expression \eqref{E:canonical_projection_q_infty} implies that the canonical projection $\ccalP_\infty^*$ is equivalent to single linkage hierarchical clustering \cite[Ch. 4]{clusteringref}. Moreover, for the case $q=\infty$, the ensuing corollary follows from Theorem~\ref{theo_unicity_q_metric}.
%
%%%  C O R O L L A RY  %%%%%%%%%%%%%%%%%%%%%%%%%%%%%%%%%%%%%%%%%
\begin{corollary}\label{cor_unicity_ultrametric}
Single linkage hierarchical clustering is the only admissible hierarchical clustering method for networks, where admissibility is given by axioms (A1)-(A2).
\end{corollary}
%%%%%%%%%%%%%%%%%%%%%%%%%%%%%%%%%%%%%%%%%

Single linkage was previously shown to posses desirable theoretical features. In \cite{clust-um}, single linkage is shown to be the only admissible method from metric spaces $\ccalM_1$ to ultrametric spaces $\ccalM_\infty$ satisfying three axioms, two of which can be derived from the axioms of Projection and Transformation here presented. This implies that, when specializing our framework for $q=\infty$, there are two main advantages between the axiomatic framework derived here and that in \cite{clust-um}: i) single linkage is the only admissible method from $\ccalN$ to $\ccalM_\infty$, and not just from $\ccalM_1$ to $\ccalM_\infty$; and ii) the third axiom considered in \cite{clust-um} is redundant, since the uniqueness result can be derived based solely on the first two.

In practice, single linkage has shown to have some undesirable features like the so-called chaining effect \cite{sibson1973slink}. Nevertheless, our construction is of utility for the practitioner who prefers other hierarchical clustering methods. More specifically, by clearly stating our desired properties as axioms, it is made clear that at least one of the axioms (A1)-(A2) must be violated when picking a method different from single linkage.

\subsection{2-metric spaces}\label{sec_2_metric_spaces}

Metric and ultrametric spaces are the two most common examples of $q$-metric spaces. However, for intermediate values of $q$, i.e., between $1$ and $\infty$, spaces with other special characteristics arise. In particular, when $q=2$, we obtain a space in which every triangle is acute. More specifically, when the distances between points satisfy the 2-triangle inequality, it can be shown that every angle in every triangle is not greater than 90 degrees. Mimicking the reasoning in Sections~\ref{Ss:metric_spaces} and \ref{Ss:ultrametric_spaces}, it follows that the canonical projection $\ccalP_2^*$ with an associated induced distance between points $x$ and $x'$ given by
\begin{equation}\label{E:canonical_projection_q_2}
d^*_2(x, x') = \min_{P_{xx'}} \sqrt{\sum_{i=0}^{l-1} W(x_i, x_{i+1})^2},
\end{equation}
is the only admissible way of inducing a structure space in a network where every resulting triangle is acute.

%!TEX root = metric_projections.tex

\section{Properties of the canonical projection}\label{S:properties}

The axioms of Projection and Transformation (A1)-(A2) uniquely determine the family of canonical projections $\ccalP^*_q$ for different $q$. Moreover, additional practical properties can be extracted from the aforementioned axioms. In this section we discuss the properties of optimality (Section~\ref{Ss:optimality}) and nestedness (Section~\ref{Ss:nestedness}).

%----------------------------------------
\subsection{Optimality}\label{Ss:optimality}

A myriad of combinatorial optimization problems exist, where the goal is to find subsets of nodes or edges of a network that are optimal in some sense. Examples of the former are graph coloring -- finding a partition of non-adjacent nodes with smallest cardinality \cite{jensen2011graph} -- and the maximum independent set problem --  finding a set of non-adjacent nodes with maximal cardinality \cite{Tarjan77maxindepset}. Examples of the latter include the traveling salesman problem -- finding a path that visits each node exactly once with smallest length \cite{Lawler85traveling} -- and the minimum bisection problem \cite{Bui87bisection} -- separating the network in two pieces with the same number of nodes so that the sum of the weights in the edges that connect the pieces is minimal. In this section we focus on this second category, where the problems are characterized by an objective function that depends on the weights of the edges of the network. 

Define the function $f:\ccalN\to\reals$ that maps every network $G$ to the minimum cost $f(G)$ of an optimization problem that depends on the structure of $G$. For the traveling salesman problem, $f(G)$ is the length of the optimal salesman's trajectory in $G$. For minimum bisection, $f(G)$ is the sum of the weights in the optimal bisection of $G$. Traveling salesman and minimum bisection are known to be NP-hard and also hard to approximate in general. This means that not only is the problem of finding optimal solutions computationally intractable, but the problem of finding approximate solutions in polynomial time is impossible as well -- impossibility unless P=NP is known for the traveling salesman problem \cite{Sahni76pcomplete} and undetermined for the minimum bisection problem \cite{Feige02polylogarithmic}. However, when the network under consideration is metric, both problems are approximable in polynomial time \cite{FernandezdelaVega04bisection, christofides1976worst}. These two examples are not isolated, there are many other combinatorial problems that are approximable when we restrict our attention to metric networks \cite{Indyk99sublinear}.

We can leverage the fact that combinatorial problems are simpler to solve in metric spaces to efficiently obtain lower bounds for $f(G)$. More specifically, we restrict our attention to cost functions $f$ that do not decrease with increasing edge weights, i.e. for networks with the same number of nodes $f(G') \geq f(G)$ if the identity map is dissimilarity reducing from $G'$ to $G$. Then, if we project a network $G$ onto a metric space $M$ where no dissimilarity is increased, we may compute the lower bound $f(M)$ efficiently. The optimal choice for this projection is the canonical map $\ccalP^*_1$ as we show next in a more general proposition for $q$-metric spaces.

%%%%%%%%%%%%%%%   F   I   G   U   R    E    %%%%%%%%%%%%%%%%%%%%%
\begin{figure}
\centering
\def \thisplotscale {0.7}
\def \unit {\thisplotscale cm}

{\large
\begin{tikzpicture}[-stealth, shorten >=2, x = 1*\unit, y=0.9*\unit]

        \node at (0,0) (1) {${\bf G}$};
        \node at (4,0) (2) {${\bf (V, d)}$};
        \node at (0,-3) (3) {${\bf (V, d^*)}$};
        \node at (4,-3) (4) {${\bf (V, d)}$};
        
        \path (1) edge [very thick, above, blue] node {$\ccalP_q$} (2);	
        \path (1) edge [very thick, left, red] node {$\ccalP_q^*$} (3);	
	\path (2) edge [very thick, right, black!50!green] node {$\ccalP_q^*$} (4);	

\end{tikzpicture}
}
\caption{Diagram of maps between spaces for the proof of Proposition~\ref{prop_canonical_projection_optimality}.}
\label{fig:relation_maps_proof_optimality}
\end{figure}
%%%%%%%%%%%%%%%%%%%%%%%%%%%%%%%%%%%%%%%%%%%%%%%%

%%%   P R O P O S I T I O N  %%%%%%%%%%%%%%%%%%%%%%%%%%%%%%%%%%%%%%%%%
\begin{proposition}\label{prop_canonical_projection_optimality}
Given an arbitrary network $G = (V, E, W)$, let $\ccalP_q:\ccalN\to\ccalM_q$ be a generic $q$-metric projection with output $(V, d) = \ccalP_q(G)$. Then, for any cost function $f$ non-decreasing in the edge weights of $G$, the canonical projection $\ccalP_q^*$ satisfies
\begin{align}\label{E:prop_canonical_projection_optimality_problem}
&\ccalP_q^* = \argmin_{\ccalP_q} f(G) - f(\ccalP_q(G)) \\
&\mathrm{s. to} \qquad d(x, x') \leq W(x, x') \quad \forall \, (x, x') \in E. \nonumber
\end{align}
\end{proposition}
%%%%%%%%%%%%%%%%%%%%%%%%%%%%%%%%%%%%%%%%%
\begin{myproof}
That $\ccalP_q^*$ is feasible, meaning that its output $q$-metric $(V, d^*) = \ccalP_q^*(G)$ satisfies the constraint in problem \eqref{E:prop_canonical_projection_optimality_problem}, can be shown using the same argument used to write expression \eqref{E:proof_theo_unicity_q_metric_050}. To see that $\ccalP_q^*$ is optimal, denote by $\ccalP_q$ a feasible $q$-metric projection with output $(V, d) = \ccalP_q(G)$. The diagram in Fig.~\ref{fig:relation_maps_proof_optimality} summarizes the relations between $G$, $(V, d)$, and $(V, d^*)$. The top (blue) and left (red) maps represent the definitions of the metric projections. The right (green) map is justified by the fact that $(V, d)$ is, by definition, a $q$-metric and that $\ccalP^*_q$ satisfies the Axiom of Projection (cf.~Prop.~\ref{prop_canonical_projection_axioms}). Moreover, notice that $d$ satisfying the constraint in \eqref{E:prop_canonical_projection_optimality_problem} guarantees that the identity map $\phi:V \to V$ from $G$ to $(V, d)$ is dissimilarity-reducing. Consequently, we combine the fact that $\ccalP_q^*$ fulfills the Axiom of Transformation (cf.~Prop.~\ref{prop_canonical_projection_axioms}) with the relations between spaces in Fig.~\ref{fig:relation_maps_proof_optimality} to write
\begin{equation}\label{E:proof_prop_canonical_projection_optimality_010}
d^*(x, x') \geq d(\phi(x), \phi(x')) = d(x, x'),
\end{equation}
for all $x, x' \in V$. Combining \eqref{E:proof_prop_canonical_projection_optimality_010} with the constraint in problem \eqref{E:prop_canonical_projection_optimality_problem}, we can write that
\begin{equation}\label{E:proof_prop_canonical_projection_optimality_020}
d(x, x') \leq d^*(x, x') \leq W(x, x'),
\end{equation}
for all $(x, x') \in E$, and optimality of $\ccalP_q^*$ follows from the non-decreasing nature of the cost function $f$.
\end{myproof}
%%%%%%%%%%%%%%%%%%%%%%%%%%%%%%%%%%%%%%%%%

The optimality result in Proposition~\ref{prop_canonical_projection_optimality} provides an efficient way to bound the minimum cost of a combinatorial optimization problem. First, an \emph{upper} bound can be achieved by finding a series of feasible solutions to the problem, e.g., particular circuits for the traveling salesman or cuts for the bisection problem, possibly aided by heuristics designed for the particular problem of interest. Second, we apply the canonical projection $\ccalP^*_1$ to the network under analysis and solve efficiently the combinatorial problem in the metric space \cite{Indyk99sublinear}, with the guarantee that obtained \emph{lower} bound is the tightest among those achievable via a metric projection that does not increase the dissimilarities.

%\red{$\ell_1$ shortest path is stable in the sense that if you modify a network slightly then a bounded modification is introduced into the projected metric spaces.}

%%%   F   I   G   U   R   E   %%%%%%%%%%%%%%%%%%%%%%%%%%%%%%%%%%%
%
\begin{figure}
\centering
\vspace{-0pt}
\def \thisplotscale {0.5}
\def \unit {\thisplotscale cm}

{\normalsize
\begin{tikzpicture}[-stealth, shorten >=2, x = 1*\unit, y=0.9*\unit]

    % Draw space blurbs
    \path (0,0) node [draw, fill = blue!10, ellipse, 
                     minimum width = 4*\unit, 
                     minimum height = 6*\unit] (networks) {} 
          ++ (-0.7,2) node {${\bf \ccalN}$};

   % \path (0,-1) node [draw, fill = blue!20, ellipse, 
     %               minimum width = 3*\unit, 
       %              minimum height = 3*\unit] (ultrametrics subset) {$\ccalM_q$};
	
	%    \path (9,-1) node [draw, fill = blue!20, ellipse, 
           %          minimum width = 3*\unit, 
             %        minimum height = 3*\unit] (ultrametrics) {$\ccalM_q$};
	
	\draw [fill = blue!20] plot [smooth cycle, tension=1] coordinates {(-1.5, -1) (1,-2.5) (0.5,-1) (0.8,2.3)};
    	\node at (0.2, 1.2) {$ {\bf \ccalM}_q$};

	\draw [fill = blue!30] plot [smooth cycle, tension=1] coordinates {(-1.3, -1) (0.45,-1.75) (0.1,-1) (0.45,0.5)};
    	\node at (-0.55, -1) {${\bf \ccalM}_{q'}$};

	\node at (6, 2.5) (N) {${\bf \ccalN}$};
	
	\node at (6, 0) (Mq) {${\bf \ccalM}_q$};
	
	\node at (6, -2.5) (Mqp) {${\bf \ccalM}_{q'}$};

    % Draw arrows
    \path[-stealth] (N) edge [thick, left, pos=0.50, red] node  {$\ccalP^*_q$} (Mq); 
    \path[-stealth] (Mq) edge [thick, left, pos=0.50, red] node  {$\ccalP^*_{q'}$} (Mqp); 
    \path[-stealth] (N) edge [thick, bend left=40, right, pos=0.50, blue] node  {$\ccalP^*_{q'}$} (Mqp);

\end{tikzpicture}

}
\caption{The canonical projection of a network in $\ccalN$ onto $\ccalM_{q'}$ (blue) is invariant to intermediate canonical projections onto spaces $\ccalM_q$ (red) for $q \leq q'$.}
\label{fig:nestedness}
\end{figure}
%%%%%%%%%%%%%%%%%%%%%%%%%%%%%%%%%%%%%%%%%%%%%%%%

%----------------------------------------
\subsection{Nestedness}\label{Ss:nestedness}

%%%%%%%%%%%%%%%   F   I   G   U   R    E    %%%%%%%%%%%%%%%%%%%%%
\begin{figure}
\centering
\def \thisplotscale {0.7}
\def \unit {\thisplotscale cm}

{\large
\begin{tikzpicture}[-stealth, shorten >=2, x = 1*\unit, y=0.9*\unit]

        \node at (0,0) (1) {${\bf G}$};
        \node at (4,0) (2) {${\bf M^0_q}$};
         \node at (8,0) (3) {${\bf M^1_{q'}}$};
        \node at (0,-3) (4) {${\bf M^1_{q'}}$};
        \node at (4,-3) (5) {${\bf M^2_{q'}}$};
        \node at (8,-3) (6) {${\bf M^1_{q'}}$};
        
        \path (1) edge [very thick, above, blue] node {$\ccalP^*_q$} (2);	
        \path (2) edge [very thick, above, dashed, red] node {} (3);	
	\path (1) edge [very thick, right, black!50!green] node {$\ccalP_{q'}^*$} (4);	
	\path (2) edge [very thick, right, black!50!green] node {$\ccalP_{q'}^*$} (5);	
	\path (3) edge [very thick, right, black!50!green] node {$\ccalP_{q'}^*$} (6);	

\end{tikzpicture}
}
\caption{Diagram of maps between spaces for the proof of Proposition~\ref{prop_nestedness}.}
\label{fig:relation_maps_proof_nestedness}
\end{figure}
%%%%%%%%%%%%%%%%%%%%%%%%%%%%%%%%%%%%%%%%%%%%%%%%

From the increasing structure imposed by the $q$-triangle inequality as $q$ increases, it follows that $\ccalM_{q'} \subseteq \ccalM_q$ for $q' \geq q$. This implies that the canonical projection of any network onto a $q'$-metric space can be alternatively achieved by a direct application of $\ccalP^*_{q'}$ or by first applying $\ccalP^*_{q}$ to the network and then applying $\ccalP^*_{q'}$ to the resulting $q$-metric space; see Fig.~\ref{fig:relation_maps_proof_nestedness}. Both approaches are equivalent, as we formally state next.

%%%   P R O P O S I T I O N  %%%%%%%%%%%%%%%%%%%%%%%%%%%%%%%%%%%%%%%%%
\begin{proposition}\label{prop_nestedness}
Given an arbitrary network $G = (V, E, W) \in \ccalN$, we have that
\begin{equation}\label{E:nestedness_010}
\ccalP^*_{q'}(G) = \ccalP^*_{q'}(\ccalP^*_q(G)),
\end{equation}
for $q' \geq q$.
\end{proposition}
%%%%%%%%%%%%%%%%%%%%%%
\begin{myproof}
Define the $q'$-metric spaces $M^1_{q'} = \ccalP^*_{q'}(G) = (V, d_1)$ and $M^2_{q'} = \ccalP^*_{q'}(\ccalP^*_q(G))= (V, d_2)$, as well as the $q$-metric space $M^0_q = \ccalP^*_q(G) = (V, d_0)$. To prove the proposition, we must show that $d_1(x, x') = d_2(x, x')$ for all $x, x' \in V$. Consider the diagram in Fig.~\ref{fig:relation_maps_proof_nestedness}, where the blue map and the two left-most green maps represent the aforementioned definitions and the remaining green map is justified by the Axiom of Projection. Notice that the proposition statement is equivalent to declaring that the red dashed map is also equal to $\ccalP^*_{q'}$. Assume that the identity map $\phi:V \to V$ is dissimilarity reducing from $G$ to $M^0_q$ and that the same identity map is also dissimilarity reducing from $M^0_q$ to $M^1_{q'}$ (top row of Fig.~\ref{fig:relation_maps_proof_nestedness}). Then, from the fact that $\ccalP^*_{q'}$ satisfies the Axiom of Transformation, it would follow that $d_1(x, x') \geq d_2(x, x') \geq d_1(x, x')$ for all $x, x' \in V$ (bottom row of Fig.~\ref{fig:relation_maps_proof_nestedness}), showing the desired equality. Thus, to complete the proof we need to show that the identity maps $\phi$ are effectively dissimilarity reducing. That $\phi$ is dissimilarity reducing from $G$ to $M^0_q$ was shown in the proof of Theorem~\ref{theo_unicity_q_metric} [cf.~\eqref{E:proof_theo_unicity_q_metric_050}]. Finally, to see that $\phi$ is dissimilarity reducing from $M^0_q$ to $M^1_{q'}$, pick an arbitrary pair of nodes $x, x' \in V$ and let $P_{xx'}=[x=x_0,\ldots, x_l=x']$ be a path achieving the minimum in \eqref{E:def_canonical_q_projection} for $d_0$ so that we can write
\begin{align}\label{E:proof_nestedness_020}
   d_0(x, x') & =  \left(\sum_{i=0}^{l-1}  W(x_i,x_{i+1})^q\right)^{1/q} \\
   & \geq \left(\sum_{i=0}^{l-1}  W(x_i,x_{i+1})^{q'}\right)^{1/{q'}} \geq d_1(x, x'), \nonumber
\end{align}
where the first inequality is obtained by combining the facts that $q' \geq q$ and the weights $W$ are non-negative, and the second inequality comes from the fact that $d_1(x, x')$ is obtained by minimizing over all paths from $x$ to $x'$ and here we are considering a particular path $P_{xx'}$. From \eqref{E:proof_nestedness_020} it follows that the identity map is dissimilarity reducing from $M^0_q$ to $M^1_{q'}$, concluding the proof.
\end{myproof}
%%%%%%%%%%%%%%%%%%%%%%%%%%%%%%%%%%%%%%%%%%%

Proposition~\ref{prop_nestedness} shows that the $q'$-metric space associated with a given network $G$ is independent of any intermediate canonical projections to $q$-metric spaces for $q \leq q'$. Intuitively, this result implies that the intermediate map $\ccalP^*_q$ induces part of the structure imposed by $\ccalP^*_{q'}$. Thus, it is equivalent to induce the whole structure in one step by applying $\ccalP^*_{q'}$ or doing it gradually by applying maps $\ccalP^*_{q}$ for $q \leq q'$.

A direct consequence of \eqref{E:nestedness_010} is that if one is interested in, e.g., computing the single linkage hierarchical clustering output of a given network $G$ (cf.~Section~\ref{Ss:ultrametric_spaces}) then there is no gain (or loss) in first projecting the network $G$ onto a metric space and then computing the clustering output of the resulting space.

%!TEX root = metric_projections.tex

\section{Search in metric structures}\label{S:search}

Given a network $G=(V, E, W)$, assume that we have access only to a subset of the network $G' = (V', E', W')$ where $V' \subset V$ and $E'$ and $W'$ are the restrictions of $E$ and $W$ to $V'$, respectively. We are then revealed an additional point $z \in V \backslash V'$ and are interested in finding the node $x \in V'$ closest to $z$, i.e., the node $x$ for which $W(z, x)$ is minimized. The described setting occurs frequently in practice, e.g., in the implementation of $k$ nearest neighbor (k-NN) methods \cite[Chapter 2]{Bishop06} where $G$ is the dataset of interest and $G'$ is the training set. The complexity of the mentioned task depends on how structured network $G$ is. When no structure is present, an exhaustive search is the only option and $z$ must be compared with every node in $V'$. By contrast, when $G$ is a metric space, then the NN of $z$ can be found efficiently by using metric trees \cite{Yianilos93, Uhlmann91, Traina00}. In this section, we propose an efficient search strategy in networks by first projecting a general network onto a metric network and then leveraging this structure for search via the construction of a metric tree.

%%%%%%%%%%%%%%%   F   I   G   U   R    E    %%%%%%%%%%%%%%%%%%%%%
\begin{figure}[t]
\centering
\def \thisplotscale {0.7}
\def \unit {\thisplotscale cm}

{\small
\begin{tikzpicture}[-stealth, shorten >=2, x = 1*\unit, y=0.9*\unit]

    % Draw bigger network
    \node [red vertex, minimum height=0.6cm, minimum width=0.6cm] at (0,  0) (1) {{Root}};
    \node [red vertex, minimum height=0.6cm, minimum width=0.6cm] at (-3, -2) (2) {L};    
    \node [red vertex, minimum height=0.6cm, minimum width=0.6cm] at (3, -2) (3) {R};
    \node at (-4, -4) (4) {$\cdots$};
    \node at (-2, -4) (5) {$\cdots$};
    \node at (2, -4) (6) {$\cdots$};
    \node at (4, -4) (7) {$\cdots$};
    
    \node [right] at (1,1) (8) {{{\footnotesize nodes: $V'$}}};
    \node [right] at (1,0.5) (8) {{{\footnotesize vantage point: $v$}}}; 
    \node [right] at (1,0) (8) {{{\footnotesize median: $\mu_v$}}}; 
    
    \node [right] at (-2,-1.5) (8) {{{\footnotesize nodes: $V_L$}}};
    \node [right] at (-2,-2) (8) {{{\footnotesize vantage point: $v_L$}}}; 
    \node [right] at (-2,-2.5) (8) {{{\footnotesize median: $\mu_{v_L}$}}}; 
    
    \node [right] at (3.8,-1.5) (8) {{{\footnotesize nodes: $V_R = V' \backslash V_L$}}};
    \node [right] at (3.8,-2) (8) {{{\footnotesize vantage point: $v_R$}}}; 
    \node [right] at (3.8,-2.5) (8) {{{\footnotesize median: $\mu_{v_R}$}}}; 

   % \path [stealth-stealth] (1) edge [right] node {{$1$}} (2);	
    \path [-stealth]  (1) edge [above left, pos=0.2] node {{{\scriptsize $\{ x \! \in \! V' \! : \! W(x, v) \! \leq \! \mu_v \}$}}} (2);
    \path [-stealth]  (1) edge [above left] node  {{}} (3);
    \path [-stealth]  (2) edge [above left] node  {{}} (4);
    \path [-stealth]  (2) edge [above left] node  {{}} (5);
    \path [-stealth]  (3) edge [above left] node  {{}} (6);
    \path [-stealth]  (3) edge [above left] node  {{}} (7);

\end{tikzpicture}
}
\caption{Vantage point tree. The whole point set $V'$ is associated with the root of the tree. A vantage point $v$ is chosen at random and $V'$ is partitioned into the left and right children of the root depending on the distance of each point to the vantage point. The process is repeated iteratively to construct the whole tree.}
\label{fig:vp_tree}
\end{figure}
%%%%%%%%%%%%%%%%%%%%%%%%%%%%%%%%%%%%%%%%%%%%%%%%

Intuitively, if $z$ is far away from a node $x$ in a metric space, i.e. $W(z, x)$ is large, then the triangle inequality implies that $z$ will also be far away from any node $x'$ close to $x$, thus, there is no need to consider nodes $x'$ as potential candidates for the NN of $z$. Metric trees formally leverage this intuition by constructing hierarchical structures of $V'$ in order to accelerate search. 
%Metric trees arise as an efficient alternative to other multidimensional space-partitioning structures, such as k-d trees \cite{Bentley75multidimensional}, which do not leverage directly the metric structure of the data but require the data to be embedded in a coordinate space. 
In this paper we focus on the vantage point tree (vp-tree) \cite{Yianilos93}, one of the most popular types of metric tree. The implementation of a metric tree is a two-step process: we first construct the tree and then utilize it for (possibly multiple) queries.

To construct a vp-tree given $G'$, we begin by associating the whole node set $V'$ to the root of the tree and we pick a node (the vantage point) at random, say $v \in V'$ . We then compute the median $\mu_v$ of the distances $W(v, x)$ from the vantage point to every other node $x \in V'$ and partition $V'$ into two blocks: one containing the nodes whose distance to $v$ is smaller than or equal to $\mu_v$ and the other one containing the rest of $V'$. The nodes in the first block are assigned to the left child of the root of the vp-tree while the right child consists of the nodes in the second block. We iteratively repeat this procedure within each of the children until every leaf in the vp-tree is associated to a single point in $V'$; see Fig.~\ref{fig:vp_tree}. For more details, see \cite{Yianilos93}. 

%%%%%%%%%% F I G U R E %%%%%%%%%%%%%%%%%%%%%%%%%
\begin{figure}[t]
\centering
\includegraphics[width=0.40\textwidth]{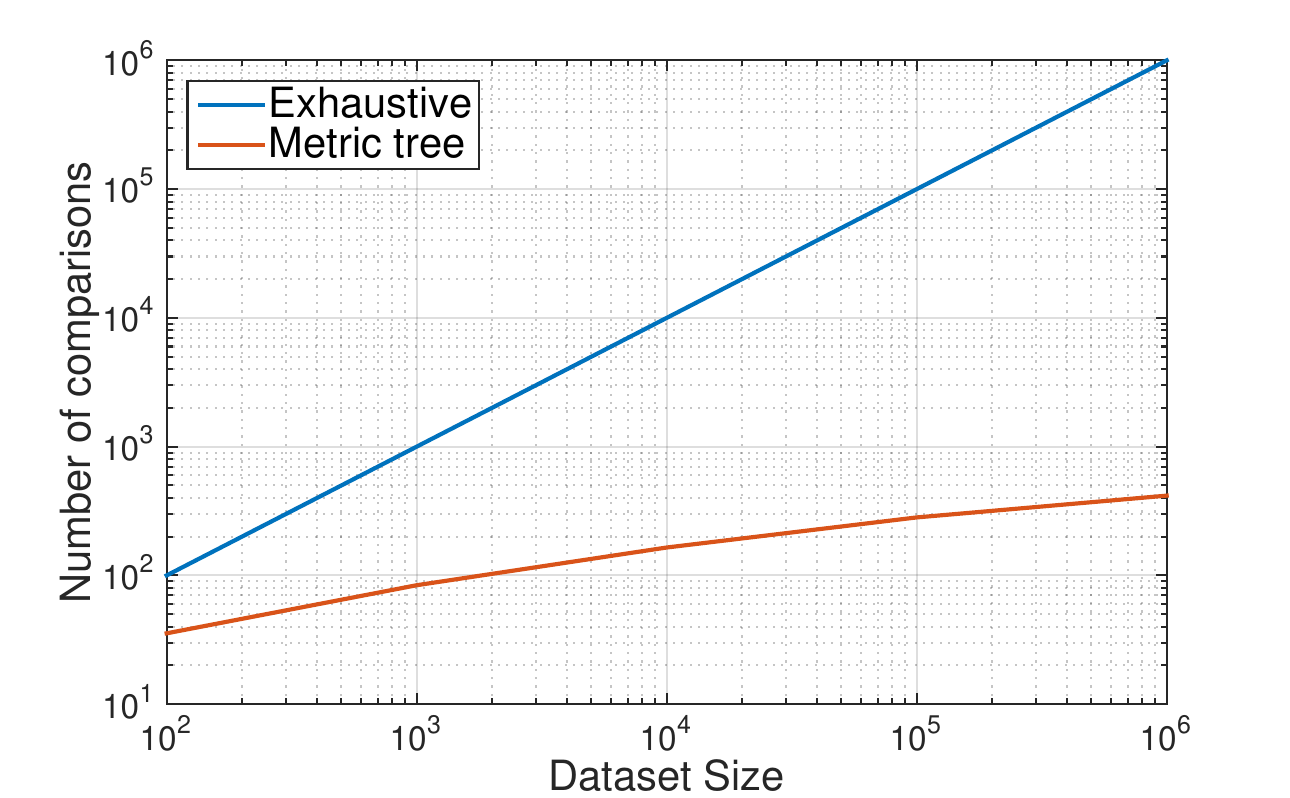}
\caption{Number of comparisons needed to find the nearest neighbor of a point in a metric space as a function of the space size for exhaustive search and search aided using metric trees.}
\label{F:graph_times}
\end{figure}
%%%%%%%%%%%%%%%%%%%%%%%%%%%%%%%%%%%%%%%%%%

To efficiently search a vp-tree for the NN of a query point $z$, we traverse the nodes of the tree and compare $z$ \emph{only with the vantage point} of the current node of the vp-tree. Moreover, we leverage the triangle inequality to discard branches of the vp-tree without even traversing them, reducing the number of measurements needed to find the NN of $z$. More specifically, assume that we are searching at an intermediate node in the vp-tree, say node $L$ in Fig.~\ref{fig:vp_tree} and the current best estimate of the NN is at distance $\tau$ from $z$, which can be initialized as $\tau = \infty$ for the root of the vp-tree. We then compute the distance $W(z, v_L)$ between $z$ and the vantage point $v_L$ associated to the current node in the vp-tree. If $W(z, v_L)< \tau$, we then update our estimate of $\tau$. In order to continue traversing the vp-tree, we follow the ensuing rules where $v$ is the vantage point of the current node in the vp-tree
\begin{align}\label{eq_rules_metric_tree}
\begin{cases}
i) \, W(z, v) \leq \mu_v - \tau  \, \Rightarrow \, \text{visit only the left child},\\
ii) \, \mu_v - \tau < W(z, v) \leq \mu_v + \tau  \, \Rightarrow \,  \text{visit left \& right child},\\
iii) \, \mu_v + \tau < W(z, v)  \, \Rightarrow \,  \text{visit only right child}.
\end{cases}
\end{align}
Even though statements $i)$ and $iii)$ in \eqref{eq_rules_metric_tree} entail that we discard part of the nodes in $V'$ during our search, the way the metric tree is constructed guarantees that the NN of $z$ is not contained among the discarded nodes. 

The construction of the vp-tree, a one-time computational effort, can be shown to have complexity $\mathcal{O}(n \log n)$ where $n$ is the cardinality of $V'$. However, once it is built it can be used to reduce the complexity of a brute force linear search from $\mathcal{O}(n)$ to an expected cost of $\mathcal{O}(\log n)$ \cite{Yianilos93}. To corroborate this, we construct metric spaces of varying sizes by embedding points in a square area of $\reals^2$ and consider their Euclidean distance as the dissimilarity values $W$. In Fig.~\ref{F:graph_times} we plot the average number of comparisons -- values of $W$ -- needed to find the nearest neighbor of a query point in this metric space as a function of $n$ for exhaustive and metric-tree search. This average is computed across 1000 queries. As expected, exhaustive search complexity grows linearly with $n$ whereas vp-tree's complexity grows logarithmically. Notice that there is a marked difference in the number of measurements required, e.g., for $n=10^6$ the metric tree search can be performed with an expected cost of 500 measurements.

%%%%%%%%%%%%%%%   F   I   G   U   R    E   :   R    A    N    D        N   E   T   W   O   R   K   %%%%%%%%%%%%%%%%%%%%%
\begin{figure*}
\centering

\begin{subfigure}{.33\textwidth}
  \centering
  \includegraphics[width=\textwidth]{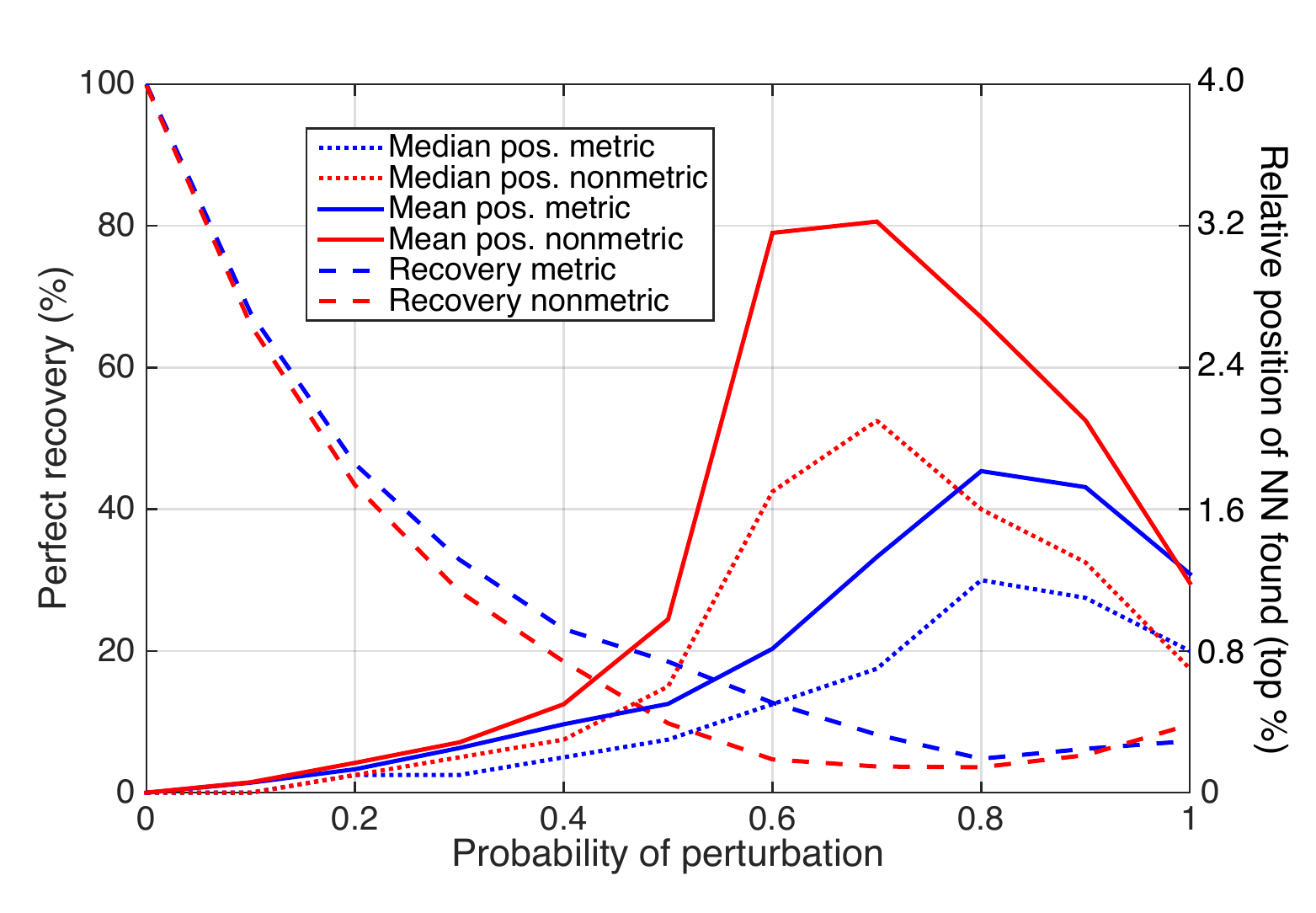}
  \caption{}
  \label{fig:search_sub1}
\end{subfigure}%
\begin{subfigure}{.33\textwidth}
  \centering
  \includegraphics[width=\textwidth]{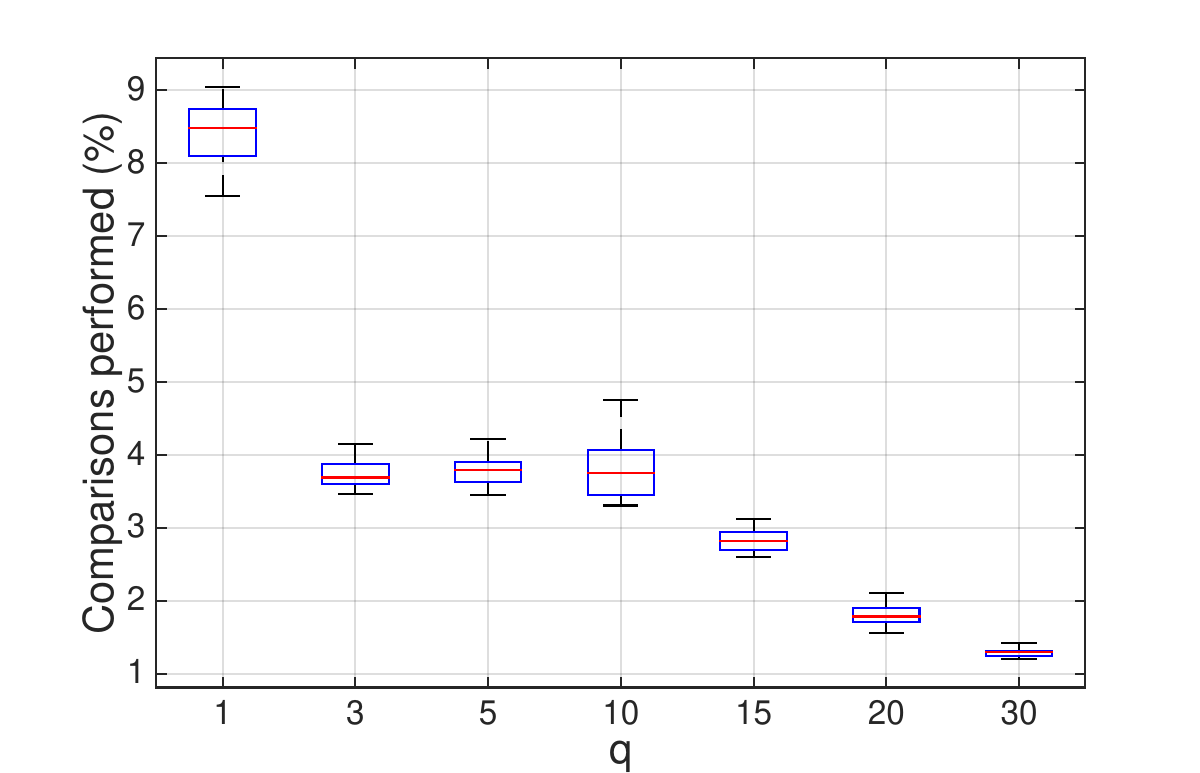}
  \caption{}
  \label{fig:search_sub2}
\end{subfigure}%
\begin{subfigure}{.33\textwidth}
  \centering
  \includegraphics[width=\textwidth]{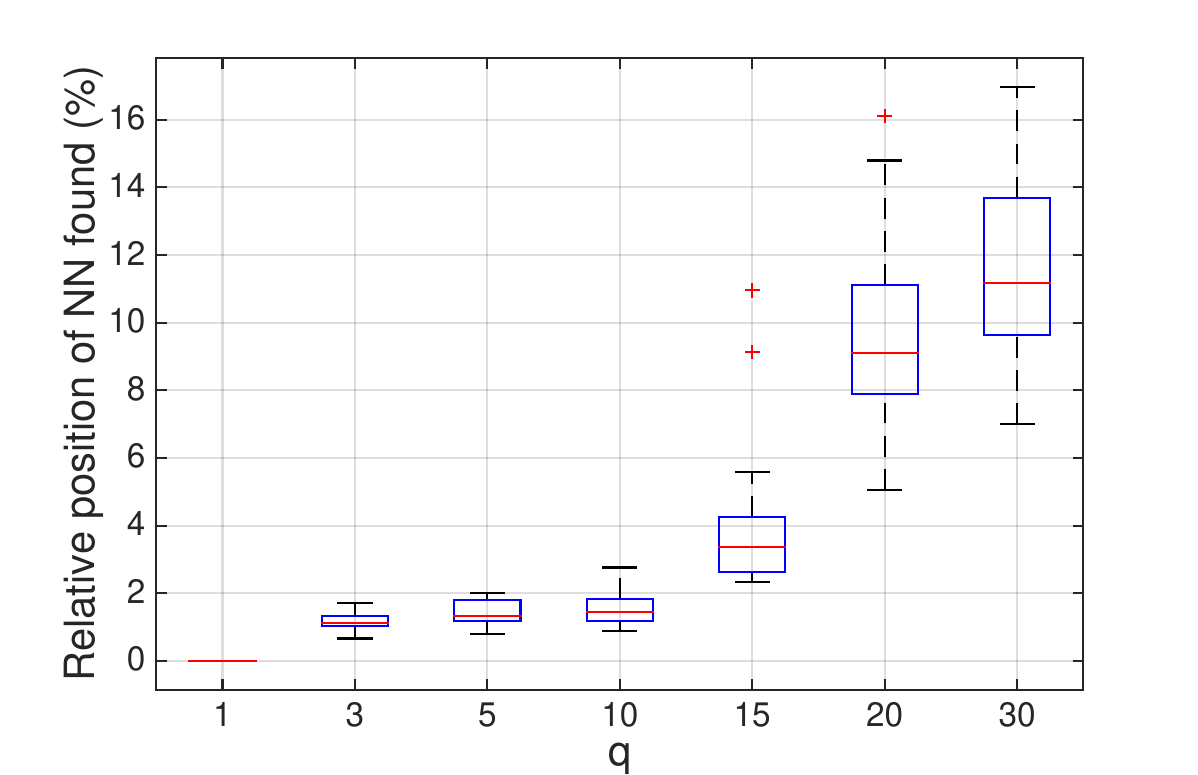}
  \caption{}
  \label{fig:search_sub3}
\end{subfigure}
\caption{(a) Percentage of perfect recovery (dashed lines, left y-axis), mean and median relative positions of search result (solid and pointed lines, right y-axis) as a function of the probability of perturbation in a metric network when the tree search is performed in the resulting non-metric space (red) and when the space is previously projected using $\ccalP^*_1$ (blue). (b) Number of comparisons needed to search a metric space when projected first onto a $q$-metric space via $\ccalP^*_q$ as a function of $q$. Larger $q$ correspond to more efficient searches. (c) Search performance as indicated by the relative position of the NN found when performing the searches in (b). Smaller $q$ correspond to more accurate searches.}
\label{fig_search_resultsexperiments}
\end{figure*}
%%%%%%%%%%%%%%%%%%%%%%%%%%%%%%%%%%%%%%%%%%%%%%%%%%%%%%%%%%%%%%%%%%%

Motivated by the computational gain depicted in Fig.~\ref{F:graph_times}, a possible way to search a non-metric network $G$ is to first project it onto a metric space $M$ via the canonical projection $M = \ccalP^*_1(G)$ and then construct a vp-tree on $M$. Notice that this construction guarantees an efficient search of the NN in $M$ of a given query. However, we are interested in finding the NN in $G$, thus, potentially committing an error. Intuitively, the furthest away the structure of $G$ is from being metric, the larger the error in the NN found. In order to illustrate this effect, we generate metric spaces obtained by randomly embedding 1000 points in $\reals^{100}$ and considering their Euclidean distances as dissimilarities between them. We then obtain (non-metric) perturbed versions of each metric space by multiplying a subset of the dissimilarities by $1 + \delta$ where $\delta$ is a random variable uniformly distributed in $[0, 10]$. The subset of dissimilarities to modify is chosen randomly with probability of perturbation $r$.
In Fig.~\ref{fig:search_sub1} we illustrate the search performance over 1000 queries as a function of $r$ (blue lines). The dashed line illustrates the percentage of perfect recovery (left y-axis), i.e., the proportion of the 1000 queries in which the node found coincides with the actual NN of the query point. The solid and the pointed lines represent, respectively, the mean and median relative positions of the actual node found (right y-axis). E.g., a value in 0\%--1\% indicates that the node found is actually contained among the 10 nearest nodes (1\% of 1000) to the query. Finally, to illustrate the value of the projection method proposed, we also illustrate the search performance when the vp-tree is constructed directly on $G$, i.e. when we apply the aforementioned construction scheme and navigation rules [cf. \eqref{eq_rules_metric_tree}] to $G$ even though it is non-metric. First of all, notice that when $r = 0$, both schemes work perfectly since $G = M$ corresponds to a metric space. For other values of $r$, the vp-tree constructed on $M$ (blue lines) consistently outperforms the one constructed on $G$ (red lines). E.g., for $r=0.6$ the median and mean relative positions of the nodes found on $M$ are in the top 0.5\% and 0.8\%, respectively, which contrast with the ones found on $G$ which are in the top 1.7\% and 3.2\%, respectively. Finally, notice that for large values of $r$ when most of the edges in $G$ are perturbed, the structure becomes more similar to a metric space and, thus, there is an improvement in the search performance both on $G$ and $M$.

When $q>1$, $q$-metric spaces contain a more stringent structure than regular metric spaces and this additional structure can be used to speed up search, as we show next.
%
%%%%% P R O P O S I T I O N %%%%%%%
\begin{proposition}\label{P:vp_tree_q_metric_spaces}
When using a vp-tree for nearest neighbor search in a $q$-metric space, the following rules ensure optimality of the node found [cf. \eqref{eq_rules_metric_tree}]
\begin{align}\label{eq_rules_metric_tree_q}
\!
\begin{cases}
i) \,W(z, v)^q \leq \mu_v^q - \tau^q \, \Rightarrow \, \text{visit only the left child},\\
ii) \,\mu_v^q - \tau^q \!<\! W(z, v)^q \!\leq\! \mu_v^q + \tau^q \, \Rightarrow \, \text{visit left \& right child},\\
iii) \, \mu_v^q + \tau^q < W(z, v)^q \, \Rightarrow \, \text{visit only right child},
\end{cases}
\end{align}
where $z$ is the query point and $v$ is the vantage point of the current node of the vp-tree.
\end{proposition}
\begin{myproof}
We need to show that by following the rules in \eqref{eq_rules_metric_tree_q} we are not discarding any node that could be the NN of our query point $z$. Assume that $i)$ is true, then the $q$-triangle inequality implies that for all $t \in V$
\begin{equation}\label{E:q_triangle_vp_tree_010}
W(v, t)^q \leq W(v, z)^q + W(z,t)^q \leq \mu_v^q - \tau^q + W(z,t)^q,
\end{equation}
where the second inequality follows from $i)$. Furthermore, for every $t$ belonging to the right child of the current node of the vp-tree we have (by construction) that $W(v, t)>\mu_v$. Combining this fact with \eqref{E:q_triangle_vp_tree_010} it follows that $W(z,t) > \tau$ for every $t$ in the right child. This implies that every node $t$ in the right child is at a distance from $z$ larger than the current best estimate $\tau$ and, thus, can be discarded. Similarly, if we assume that $iii)$ is true, we may leverage the $q$-triangle inequality to write that for all $t \in V$
\begin{equation}\label{E:q_triangle_vp_tree_020}
\mu_v^q + \tau^q < W(z, v)^q \leq W(z, t)^q + W(t,v)^q.
\end{equation}
If we combine \eqref{E:q_triangle_vp_tree_020} with the fact that for every $t$ in the left child $W(t, v) \leq \mu_v$, it follows that $W(z,t) > \tau$ for every $t$ in the left child, concluding the proof.
\end{myproof}
%%%%%%%%%%%%%%%%%%%%%%%%

Notice that when $q=1$, the conditions in \eqref{eq_rules_metric_tree_q} boil down to those in \eqref{eq_rules_metric_tree}, as expected. Further, the range encompassed by \eqref{eq_rules_metric_tree_q}-$ii)$ is smaller than that in \eqref{eq_rules_metric_tree}-$ii)$, and decreases in size with larger $q$. To see this, compute the $q$-th of \eqref{eq_rules_metric_tree}-$ii)$ and, from the fact that $\mu_v$ and $\tau$ are positive, it follows that $\mu_v^q + \tau^q \leq (\mu_v + \tau)^q$ showing that the upper bound in \eqref{eq_rules_metric_tree_q}-$ii)$ is tighter. Similarly, whenever $0 \leq \tau \leq \mu_v$ -- otherwise, both lower bounds are effectively zero --, we have that $ (\mu_v - \tau)^q \leq \mu_v^q - \tau^q$, showing that the lower bound in \eqref{eq_rules_metric_tree_q}-$ii)$ is tighter as well compared to the one in \eqref{eq_rules_metric_tree}-$ii)$. This implies that in more structured spaces -- larger $q$ -- it is more likely to discard parts of the vp-tree -- satisfying conditions $i)$ or $iii)$ -- speeding up the search. Based on this observation, one can project a metric space onto $q$-metric spaces with $q>1$ in order to increase search speed with the cost of decreasing search performance due to the deformation introduced when projecting the original metric space. Figs.~\ref{fig:search_sub2} and \ref{fig:search_sub3} illustrate this tradeoff. We generate 20 metric spaces with 1000 points each and perform 100 queries in each metric space. In Fig.~\ref{fig:search_sub2} we illustrate the distribution of the 20 averages of the number of comparisons needed (as a percentage of 1000) to perform the search when first projecting the metric space applying $\ccalP^*_q$ for varying $q$. In Fig.~\ref{fig:search_sub3} we plot the relative position of the NN found as a function of $q$. Notice that when $q=1$, the node found is always the correct one (0\%) since the original space was chosen metric. Furthermore, when, e.g., $q=3$ the number of measurements needed to search the vp-tree is reduced to around half of those needed for metric spaces (cf.~Fig.~\ref{fig:search_sub2}) while the neighbors found are within the top 2\% candidates among the 1000 points (cf.~Fig.~\ref{fig:search_sub3}). For large values of $q$, the reduction in computation is noticeable -- around 8 times for $q=30$ -- but the detriment in performance is also large. Depending on the application, the value of $q$ can be tuned to find the correct equilibrium between computational efficiency and search performance.

%!TEX root = metric_projections.tex

%%%%%%%%%%%%%%%%%%%%%%%%%%%%%%%%%%%%%%%%%%%%%%%%%%%%%%%%%%%%%%%
\section{Conclusion}\label{S:conclusion}
%%%%%%%%%%%%%%%%%%%%%%%%%%%%%%%%%%%%%%%%%%%%%%%%%%%%%%%%%%%%%%%

We analyzed how to project networks onto $q$-metric spaces, a generalization of metric spaces that encompasses a larger class of structured representations. We defined the axioms of Projection and Transformation as desirable properties of the projection methods and showed that for each $q$-metric space there is a unique canonical way of projecting any network onto it. In particular, this axiomatic framework provided theoretical support for the computation of shortest path distances between nodes as well as for single linkage hierarchical clustering of networked datasets. Moreover, we showed that $q$-metric projections can be used in practice to approximate combinatorial optimization problems in graphs and to efficiently search a network for the nearest neighbor of a given query point.

%%%%%%%%%%%%%%%%%%%%%%%%%%%%%%%%%%%%%%%%%%%%%%%%%%%%%%%%%%%%%%%
%\section{Applications}
%%%%%%%%%%%%%%%%%%%%%%%%%%%%%%%%%%%%%%%%%%%%%%%%%%%%%%%%%%%%%%%

%----------------------------------------
%\subsection{Search}

%----------------------------------------
%\subsection{Graph Theoretical Problems}

%%%%%%%%%%%%%%%%%%%%%%%%%%%%%%%%%%%%%%%%%%%%%%%%%%%%%%%%%%%%%%%
%\section{Conclusion}
%%%%%%%%%%%%%%%%%%%%%%%%%%%%%%%%%%%%%%%%%%%%%%%%%%%%%%%%%%%%%%%

%%%%%%%%%%%%%%%%%%%%%%%%%%%%%%%%%% R E F E R E N C E S %%%%%%%%%%%%%%%%%%%%%%%%
\urlstyle{same}
\bibliographystyle{IEEEtran}
\bibliography{metrics_biblio.bib}
%%%%%%%%%%%%%%%%%%%%%%%%%%%%%%%%%%%%%%%%%%%%%%%%%%%%%%%%%%%%%%%%%%%%%

\end{document}